\documentclass[aps,onecolumn,preprintnumbers,nofootinbib,superscriptaddress,11pt]{revtex4}
\usepackage{amsmath} \usepackage{graphicx} \usepackage{amsfonts}
\usepackage{array} \usepackage{amsthm} \usepackage{bm}
\usepackage{palatino} \usepackage{mathpazo} %\usepackage{mathptmx}
\usepackage{supertabular}

\usepackage[breaklinks]{hyperref}
\usepackage{color}

\usepackage{epsfig}
\usepackage{epstopdf}
\usepackage{graphicx}

\newcommand{\be}{\begin{equation}}
\newcommand{\ee}{\end{equation}}
\newcommand{\ba}{\begin{eqnarray}}
\newcommand{\ea}{\end{eqnarray}}
\newcommand{\bal}{\begin{align}}
\newcommand{\eal}{\end{align}}
\newcommand{\lb}{\label}

\newcommand{\bw}{\begin{widetext}}
\newcommand{\ew}{\end{widetext}}

\begin{document}

\title{Accretion on High Derivative Asymptotically Safe Black Holes}

\author{M. Umar Farooq}\email{m\textunderscore ufarooq@yahoo.com}
\affiliation{DBS and H, College of E and ME, National University of Sciences and Technology (NUST), H-12, Islamabad, Pakistan}

\author{Ayyesha K. Ahmed}\email{ayyesha.kanwal@sns.nust.edu.pk}
\affiliation{Department of Mathematics, School of Natural
    Sciences (SNS), National University of Sciences and Technology
    (NUST), H-12, Islamabad, Pakistan}

\author{Rong-Jia Yang}\email{yangrongjia@tsinghua.org.cn}
\affiliation{College of Physical Science and Technology, Hebei University, Baoding 071002, China}

\author{Mubasher Jamil}\email{mjamil@zjut.edu.cn}
\affiliation{ Institute for Theoretical Physics and Cosmology, Zhejiang University of Technology, Hangzhou 310023, China}
\affiliation{Department of Mathematics, School of Natural Sciences (SNS), National University of Sciences and Technology (NUST), H-12, Islamabad, Pakistan}
\affiliation{Canadian Quantum Research Center, 204-3002, 32 Ave, Vernon, BC, V1T 2L7, Canada}

\begin{abstract}
Asymptotically safe gravity is one effective approach to quantum gravity. It is
important to differentiate the modified gravity inspired by asymptotically safe gravity. In this paper, we examine the matter particles dynamics near the improved version of Schwarzschild black hole. We assume that in the context of asymptotically safe gravity scenario the ambient matter surrounding the black hole is of isothermal in nature and investigate the spherical accretion of matter by deriving solutions at critical points. The analysis for the various values of the state parameter for isothermal test fluids, viz., $k=1,~1/2,~1/3,~1/4$ show the possibility of accretion onto asymptotically safe black hole. We formulate the accretion problem as Hamiltonian dynamical system and explain its phase flow in detail which reveals interesting results in asymptotically safe gravity theory.

\end{abstract}

\maketitle
\section{Introduction}
The process of accretion of matter onto a black hole is one of the interesting topics in theoretical physics. This process is mainly responsible for the formation of Quasi-periodic oscillations and emission of gravitational waves. The accretion phenomena signifies its role in the formation of astronomical objects, such as stars, planets, galaxies, quasars etc. The most important phenomena in the universe such as gamma-ray burst, X-ray binaries, active glactic nuclei, tidal disruption events are mainly due to the accretion of gas onto the black holes. Accretion disk forms when the gaseous matter rotates and accumulates around the black hole. In the last century, it was realized that it is the gravity which powers most of the luminous objects in the universe through accretion. Our main aim is to study accretion onto black hole that are known to be the strongest gravitating objects which are responsible to emit high energy flux from the astronomical objects and also they have an event horizon which acts as a border through which fluid can moves inside.\\
The history of the accretion process in the Newtonian gravity began in $1952$ by Bondi \cite{Bondi} and later on by Michel \cite{Michel} in $1972$ in the context of general relativity. The difference between the Bondi and the relativistic accretion models is that the former allows stellar winds or ejecta $(v>0)$ from the stellar surface which is the opposite of the accretion $(v<0)$, whereas the same process of ejecta does not straight forwardly applies to black holes since black holes are not composed of gas and has no gaseous surface. Any kind of ejection such as jets from the black holes occur only in the presence of charged plasma floating around black hole under the effect of strong magnetic fields. We have ignored these considerations in this paper. Following the Michel approach the discussion of critical points in relation to accretion was discussed in \cite{Begelman}. A detail study of accretion process onto spherically symmetric black holes in general relativity and other theories of gravities can be found in \cite{Pet,Mob,Gid,Sha,John,Deb,Gang,Babi,Jim,Bha,Mach,Machh,Kar,Yang}, and the references wherein. Babichev et. al., \cite{Babichev1,Babichev2} have shown that the phantom accretion onto a black hole decreases its mass. On the other hand, if the black hole solutions are considered in the Friedmann-Robertson-Walker universe, it was observed that the black hole accretion may increase the mass of gravitational object \cite{Gao}.\\
Black holes that stand as a fundamental part of our universe are interestingly the most intriguing solutions of Einstein field equations. Einstein general theory of relativity explains very well the exterior and horizons of black holes but fails to describe the physics of deep central region which is effected strongly by the quantum effects. In this context, Weinberg \cite{Weinberg} proposed a theory of asymptotic safe gravity (ASG) which embed gravity in the quantum field theory framework. As its central property, the effective average action satisfies a formally exact functional renormalization group equation, which by now has accumulated substantial evidence that the gravitational renormalization group flow possesses a nontrivial fixed point which could provide the ultra-violet completion of gravity at trans-Planckian energies.
%Asymptotically safe gravity (ASG) theory was proposed by  which is the effective quantum field description of a gravitation theory, may be UV complete and perturbatively renormalizable. \\
%
%Furthermore, Shapiro and Teukolsky \cite{Shapiro}, Babichev \cite{Babichev} and Debnath \cite{Debnath} have contributed a lot towards the literature of accretion process. Recently, quantum gravity corrections to accretion onto a Schwarzschild black hole was discussed in \cite{Y1}, accretion onto a Kiselev black hole was considered in \cite{Y2}, an analytic solution for accretion of a gaseous medium with an equation of state ($P=\rho$) onto a Reissner-Nordstrom black hole was obtained in \cite{Y3}, effects of Lorentz breaking on the accretion was investigated in \cite{Y5}. More interesting, constraints on the the ratio of mass to charge was derived from the accretion onto a Tangherlini-Reissner-Nordstrom black hole \cite{Y4}. Exact solutions were obtained for dust shells accreting towards a
%black hole in \cite{Liu,Zhao}. Moreover in the series of recent papers \cite{A1,A2,A3,t1,1t,PMach,Mjamil}, accretion of a spherically symmetric spacetime is investigated in detail.
The interesting idea of Asymptotically safe gravitational theory has been applied to theories of gravity (such as Einstein gravity \cite{R} and $f(R)$ gravity \cite{Ohta}) and to cosmology \cite{Litim,Yangg} and black hole \cite{Bonanno}. In this paper we study relativistic accretion problem in the context of infra-red limit of asymptotically safe scenario. We follow the Hamiltonian dynamical formalism on the phase space $(r,n)$ where $r$ is the areal radius and $n$ is the particle density of the fluid. We assume that the improved version of Schwarzschild black hole is surrounded by a special type of perfect fluid namely the isothermal and investigate the matter particle dynamics by deriving solutions at critical points. The solutions we obtain for the accretion problem describe the Michel flow and the critical point through which it passes. So to discuss critical flows, we investigate the effect of ASG on the accretion process and is the main purpose of this work.
Expressing the accreting matter by the isothermal equation of state we give a complete description of the fluid flow behavior near the black hole. \\
The remaining part of this paper is structured as: in section II we first give a brief review of field equations to define the static, spherically symmetric black hoe metric in ASG within Infra-Red limit. We also present some fundamental equations related to accretion to explain a steady-state, radial perfect fluid flow by specifying our assumptions on the fluid equation of state and enunciate our results. In section III, governing equations for improved Schwarzschild black hole accretion and conservation laws are presented. We then evalute our results at sonic points by considering the isothermal fluid and analyze its flow by choosing suitable values of state parameter. We also formulate the fluid equations as two-dimensional Hamiltonian dynamical system on the phase space $(r,n)$ by assuming that the Hamiltonian depends on the accretion rate in section IV. We perform a detailed analysis by providing numerical plots for the phase flow of isothermal fluid on modified Schwarzschild background and discuss the effect of coupling parameter. Conclusions are presented in the last section.
\section{Notation and equations for spherical accretion in ASG with in Infra-Red limit}
Recently, Cai and Easson \cite{cai} found black hole solution in ASG scenario considering higher derivative terms in their investigation. They discuss how the inclusion of quantum corrections modifies the Schwarzschild black hole solution.
So according to \cite{cai} the geometry of a static spherically symmetric Schwarzschild (anti)-de Sitter black hole in ASG in the $IR$ limit is given by
\begin{eqnarray}\label{1}
ds^{2}&=&-\Big(1-\frac{2GM}{r}+\frac{2G^2M\xi}{r^3}\Big)dt^{2}+\Big(1-\frac{2GM}{r}+\frac{2G^2M\xi}{r^3}\Big)^{-1}dr^{2}+r^2d\theta^{2}+r^2sin^{2}\theta d\phi^{2},
\end{eqnarray}
where $G$ and $M$ denotes the gravitational constant and mass of the black hole respectively. The outer horizon which is nothing but the null hypersurface of the modified version (\ref{1}) of Schwarzschild black hole taking quantum corrections into account can be written in approximate form as
\begin{eqnarray}\label{rh}
r_{h}&=&\frac{2GM}{3}\Big[1-2\cosh\Big(\frac{1}{3}\cosh^{-1}\beta\Big)\Big],
\end{eqnarray}
where $\beta=\frac{27\xi}{8GM^2}-1$. Note that $r_h$ given in (\ref{rh}) is the only real root of $1-\frac{2GM}{r}+\frac{2G^2M\xi}{r^3}=0$, which can be calculated by using Weierstrass Polynomial $r=z+\frac{2GM}{3}$. Expanding Eq. (\ref{rh}) to the leading order of $\xi$ we can approximate it as \cite{cai}
\begin{eqnarray}\label{IR}
r_{IR}\simeq2GM-\frac{\xi}{2M}.
\end{eqnarray}
It can be seen that if we set the running coupling parameter $\xi=0$ in (\ref{1}) and (\ref{IR}), we can retrieve respectively the classical Schwrazschild black hole metric and the corresponding event horizon. Here we review some important equations describing the steady state Michel flow on a Schwrazschild (anti) de Sitter black hole in ASG. For more details and generalization to even more general static spherically symmetric black hole background, the reader is suggested to see the references \cite{Kar,Cha1,Cha2}.\\
As described in the introduction, we model the flow by a relativistic perfect fluid, thereby neglecting the effects related to viscosity or heat transport and further assume that the fluid's energy density is sufficiently small so that its self-gravity can be neglected. Now we assume that the flow of the perfect fluid onto the improved Schwrazschild black hole is steady-state flowing in the radial direction described by the particle density $n$ (also called baryonic number density), pressure $p$ and the energy density $e$ by an observer moving along the fluid four-velocity $u^\alpha u_\alpha=-1$. To investigate the accretion process onto high derivative black hole as described above we need to review the fundamental equations of accretion for the underlying geometry of spacetime.\\
The accretion dynamics of a perfect matter is governed by the following conservation laws
\begin{equation}\label{moom}
\nabla_{\alpha}J^\alpha=0,
\end{equation}
\begin{equation}\label{momm}
\nabla_{\alpha}T^{\alpha\beta}=0,
\end{equation}
where $J^{\alpha}=nu^{\alpha}$ is the particle current density and $T^{\alpha  \beta}=nhu^{\alpha}u^{\beta}+pg^{\alpha\beta}$ is the stress energy tensor and $\nabla$ refers to the covariant derivative with respect to the spacetime metric.  Here and onward we assume that $h$ denotes the enthalpy per particle defined by $h=\frac{p+e}{n}$ \cite{Ficek} where $h=h(n)$ is a function of the particle density $n$ only. In the spherical symmetry stationary case the above equations (\ref{moom}) and (\ref{momm}) reduce to
\begin{equation}\label{M}
r^2nu=const=K,
\end{equation}
\begin{equation}\label{L}
h\Big(1-\frac{2GM}{r}+\frac{2G^2M\xi}{r^3}+u^{2}\Big)^{1/2}=const=L,
\end{equation}
which expresses the conservation of particle and energy flux through a sphere of constant areal radius $r$.
%and if we also assume that the fluid is flowing smoothly then we can integrate the Eq. (\ref{eng}) and (\ref{mom}) which yields
%\begin{eqnarray}
%r^{2}n u&=&C_{1},\label{17}\\
%h\Big(1-\frac{2GM}{r}+\frac{2G^2M\xi}{r^3}+u^{2}\Big)^{1/2}&=&C_{2},\label{18}
%\end{eqnarray}
%where $L$ and $M$ are constants.
We stress here that to analyze the perfect fluid flow, Eqs. (\ref{M}) and (\ref{L}) will play main role in the background of improved Schawrzschild black hole as they will be helpful to convert the present problem into Hamiltonian dynamical system.
\section{Flow Behavior at Critical Point}
Physically, a critical point $r=r_c$ describes the transition of the flow's radial velocity measured by static observer from subsonic to supersonic. If we consider the barotropic fluid for which there exists a constant pressure throughout (i.e. $h=h(n)$) then its equation of state can be expressed as \cite{Ficek}
\begin{eqnarray}\label{33}
\frac{dh}{h}=a^2 \frac{d n}{n},
\end{eqnarray}
where $a$ denotes the local speed of sound.

Also on differentiating Eq. (\ref{M}) and (\ref{L}) with respect to $r$ we obtain
\begin{eqnarray}\label{new1}
\frac{du}{dr}&=&\frac{2u}{r}~.~\frac{c_{s}^{2}\Big(1-\frac{2GM}{r}+\frac{2G^2M\xi}{r^3}+u^2\Big)-\frac{GM}{2r}-\frac{3G^2M\xi}{2r^3}}{u^2-c_{s}^{2}\Big(1-\frac{2GM}{r}+\frac{2G^2M\xi}{r^3}+u^2\Big)},
\end{eqnarray}
where $c_{s}^{2}=k$ is the square of the speed of sound and $k$ is a state parameter for the isothermal equation of state (EoS) $p=ke$. The above equation (\ref{new1}) can be converted into 2-dimensional autonomous Hamiltonian dynamical system as follows
\begin{eqnarray}
f_{1}(r,u)&=&\frac{dr}{dl}=r\Big\{u^2-c_{s}^{2}\Big(1-\frac{2GM}{r}+\frac{2G^2M\xi}{r^3}+u^2\Big)\Big\},\label{new2}\\
f_{2}(r,u)&=&\frac{du}{dl}=2u\Big\{c_{s}^{2}\Big(1-\frac{2GM}{r}+\frac{2G^2M\xi}{r^3}+u^2\Big)-\frac{GM}{2r}-\frac{3G^2M\xi}{2r^3}\Big\},\label{new3}
\end{eqnarray}
with arbitrary parameter $l$, whose phase portraits consist of $r$ verses $s$ indicate solutions of (\ref{M}) and (\ref{L}). In order to obtain critical points we put the right hand sides of  (\ref{new2}) and (\ref{new3}) equal zero which after solving yield
%\begin{eqnarray}
%u_{\ast}^2-c_{s\ast}^{2}\Big(1-\frac{2GM}{r_{\ast}}+\frac{2G^2M\xi}{r_{\ast}^3}+u_{\ast}^2\Big)&=&0,\label{new4}\\
%c_{s\ast}^{2}\Big(1-\frac{2GM}{r_{\ast}}+\frac{2G^2M\xi}{r_{\ast}^3}+u_{\ast}^2\Big)-\frac{GM}{2r_{\ast}}-\frac{3G^2M\xi}{2r_{\ast}^3}&=&0.\label{new5}
%\end{eqnarray}
%To find the critical point and speed of the moving fluid one has to define the equation of state. However, solving the above system (\ref{new5}),(\ref{new6}), one can obtain
\begin{eqnarray}\label{new6}
u_{c}^{2}&=&\frac{GM}{2r_{c}}+\frac{3G^{2}M\xi}{2r_{c}^{3}},
\end{eqnarray}
\begin{eqnarray}\label{new7}
c_{s}^{2}&=&\frac{\frac{GM}{2r_{c}}+\frac{3G^{2}M\xi}{2r_{c}^{3}}}{1-\frac{3GM}{2r_{c}^{2}}+\frac{7G^{2}M\xi}{2r_{c}^{3}}},
\end{eqnarray}
By using (\ref{new6}) and (\ref{new7}) we can get the sonic points. Here the sonic points refer to the critical points of the dynamical systems (\ref{new2}) and (\ref{new3}).
%gives critical radius and we can obtain $u_\ast$ from (18) the critical point for various types of fluid matter. We point out here that an equivalent pair of expressions to obtain critical radius and speed
%\begin{eqnarray}
%(a_\ast^2=k)\Big[\frac{1}{4}r_\ast f_{\ast,r_\ast}+f_\ast\Big]=\Big[\frac{1}{4}r_\ast f_{\ast,r_\ast}\Big]\\
%(u_\ast^r)^2=\frac{1}{4}r_\ast f_{\ast,r_\ast}.\label{cp}
%\end{eqnarray}
% which combinedly give the exact location of sonic point was derived by two of us with other colleagues [32,33,34????]\\
From Eqs. (\ref{M}) and (\ref{L}) after performing some intermediate steps, one can arrive at very important equation which will be helpful to describe the critical flows of the fluid under consideration
\begin{eqnarray}\lb{43}
\sqrt{1-\frac{2GM}{r}+\frac{2G^2M\xi}{r^3} +(u)^2}= Ar^{2k}u^{k}.
\end{eqnarray}
%We shall use these equations to determine the critical point $(r_\ast,u_\ast)$ which will help us to analyze the nature of moving fluid.\\
For standard equation of state, the critical point $(r_{c},u_{c})$ is the saddle point and so the solution must pass through this saddle critical point. The detailed discussion of this critical point will be presented in the forthcoming sections.
\section{Hamiltonian Analysis for Isothermal Test Fluids}
In order to analyze the perfect matter flow, it is useful to utilize a dynamical system whose orbits consist of graphs of solutions of the system (\ref{M}) and (\ref{L}). Such a system can be defined conveniently in terms of $r$ and $v$ (where $v$ is the three-velocity of the fluid). Now we formulate our problem in terms of Hamiltonian dynamical system on the phase space $(r,n)$, where the vector field describing the dynamics is the Hamiltonian vector field associated with the function $F(r,n)$. By assumption, $F$ is constant along the trajectories of phase flow and thus meets the definition of level curves. The main usefulness of converting the accretion problem into dynamical system is that the fluid behavior near the critical point of $F$ can be analyzed using standard tools of theory of dynamical systems \cite{Har,Per}.\\
In Eqs. (6) and (7) we have used two integrals of motion $K$ and $L$. We stress here that any one of them , or any combination of these integrals can be utilized as Hamiltonian system for the fluid flow. Let us assume that the Hamiltonian system as function of two variables $r$ and $v$ and is square of the left hand side of (\ref{L}), i.e.,
\begin{equation}
\mathcal{H}(r,v) =h^2\Big(1-\frac{2GM}{r}+\frac{2G^2M\xi}{r^3}+u^{2}\Big)
\end{equation}
which in more general form can be written as
\begin{equation}
\mathcal{H}(r,v) =\frac{h^2(r,v)(f(r)+u^2)}{1-v^2}.
\end{equation}
Let us introduce the following pair of dynamical system
\begin{equation}
\dot{r}=\mathcal{H}_{,v}~~~~~~~\dot{v}=-\mathcal{H}_{,r}~,
\end{equation}
where the over dots denote the $\tilde{t}$ derivatives and the $\mathcal{H}_{,r}$ and ~$\mathcal{H}_{,v}$ denote partial derivatives of $\mathcal{H}$ with respect to $r$ and $v$ respectively. By evaluating the right hand side of above equation and then equating to zero results in the desired critical point $(r_c, u_c)$. Therefore by doing so, we obtain the following fundamental pair of equations
\begin{equation}
v_c^2=a_c^2,~~~~~~r_c(1-a_c^2)f_c,r_c=4f_ca_c^2,
\end{equation}
which are thus helpful to derive the following important equations
\begin{eqnarray}
(a_c^2=k)\Big[\frac{1}{4}r_c f_{c,r_c}+f_\ast\Big]=\Big[\frac{1}{4}r_c f_{c,r_c}\Big]\\
(u_c^r)^2=\frac{1}{4}r_c f_{c,r_c}.\label{cp}
\end{eqnarray}
We point out here that the above pair of Eqs. (19) and(20) is equivalent to (12) and (13) and provide the critical radius and critical speed of the moving fluid. So we shall use these equations to locate the position of the critical point $(r_\ast, u_\ast^r)$. We know that one of the most appreciative tool for energy conservation is Hamiltonian. In the current study, we use the precise form of the general Hamiltonian (15) in the variables $r$ and $v$ for the isothermal test fluid can be expressed as
\begin{eqnarray}\label{41b}
\mathcal{H}(r,v) &=& \frac{\Big(1-\frac{2GM}{r}+\frac{2G^2M\xi}{r^3}\Big)^{1-k}}{(1-v^{2})^{1-k}v^{2k}r^{4k}},
\end{eqnarray}
where $k$ is the state parameter and $v$ is ordinary 3-dimensional speed of the fluid which is given by
\begin{eqnarray}\label{41bb}
v^{2}=\frac{u^{2}}{1-\frac{2GM}{r}+\frac{2G^2M\xi}{r^3}+u^{2}}.
\end{eqnarray}
We remark here that $u$ is well defined everywhere and the velocity $v$ is defined outside the horizon. For complete derivation of these fundamental equations, interested reader is suggested to see the papers \cite{A1,A2,A3}.
\subsection{Isothermal Fluids}
Expressing the accreting matter by the isothermal equation of state (EoS) $P=ke$ (where $k$ is a state parameter), we present a complete description of the fluid flow behavior near the black hole.\\
\begin{figure}[!ht]
	\centering
	\minipage{0.45\textwidth}
	\includegraphics[width=5.9cm,height=5.4cm]{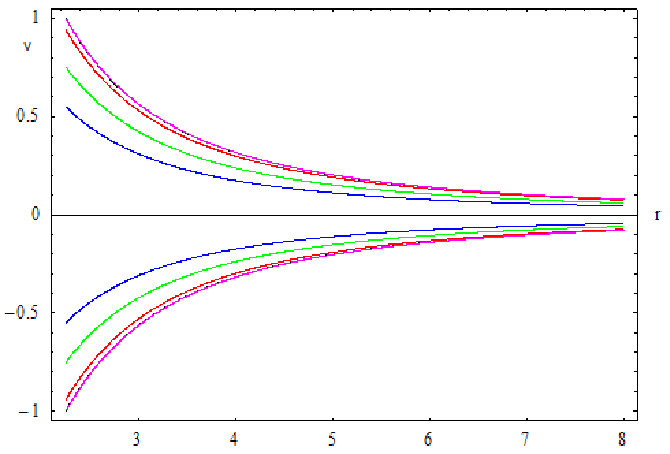}
	\label{f1}
	\endminipage\hfill
	\minipage{0.45\textwidth}
	\includegraphics[width=5.9cm,height=5.4cm]{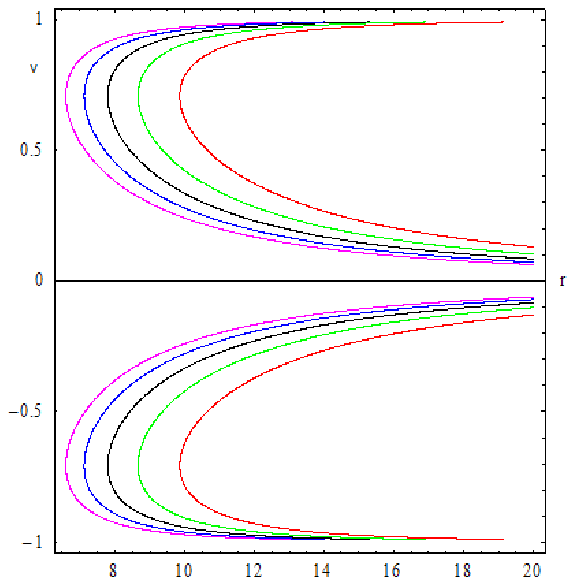}
	\label{f2}
	\endminipage\hfill\\
	\minipage{0.45\textwidth}
	\includegraphics[width=6.4cm,height=5.4cm]{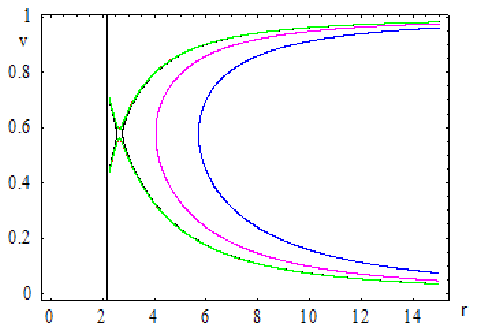}
	\label{f3}
	\endminipage\hfill
	\minipage{0.45\textwidth}
	\includegraphics[width=6.5cm,height=5.4cm]{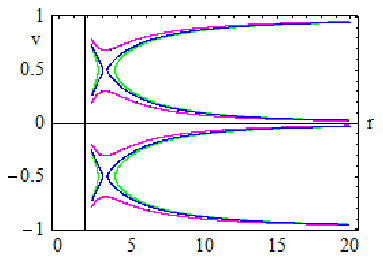}
	\label{f4}
	\endminipage\hfill
	\caption{Contour plot of Hamiltonian $\mathcal{H}(\ref{41b})$ for ultra-stiff $(k=1)$, ultra-relativistic $(k=1/2)$, radiation $(k=1/3)$ and sub-relativistic $(k=1/4)$ fluids where $M=1, G, \xi=0.5$. The black curve in these graph is the curve that pass through the saddle critical point i.e. $\mathcal{H}=\mathcal{H}_{\ast}$}\label{f5}
\end{figure}
\\
i)~~If the isotropic pressure and the energy density of the fluid particles is same then it is called ultra stiff fluid. In this case state parameter has the value $k=1$. This value of state parameter reduces the expression (19) to $f_c=0$ which yields $r_c=r_h$, i.e., the critical radius and event horizon coincides. In this case the Hamiltonian (21) reduces to
\begin{eqnarray}\label{46a}
\mathcal{H}&=&\frac{1}{v^{2}r^{4}}.
\end{eqnarray}
As above Hamiltonian shows constant of motion i.e., $\mathcal{H}=\mathcal{H}_0$ and we observe that $v$ behaves as $\frac{1}{r}$. Now to explain the physical behavior of the fluid flow we need to sketch contour plots of $\mathcal{H}(r_c,v_c)=\mathcal{H}_c$. From the figure 1 on top left, the black curve shows the solution for $\mathcal{H}=\mathcal{H}_{c}$, the red curve shows the solution for $\mathcal{H}=\mathcal{H}_{c}+0.005$, the green curve shows the solution for $\mathcal{H}=\mathcal{H}_{c}+0.02999$, the magenta curve is for $\mathcal{H}=\mathcal{H}_{c}-0.0001$ and the blue curve is for $\mathcal{H}=\mathcal{H}_{c}-0.09$. In conclusion, we observe that for $v>0$ there is particle emission and for $v<0$ it depicts the fluid accretion.\\ %(Purely supersonic accretion or flow out along the blue, black and magenta curve).
%The plot shows the trajectories of solutions to Eq. (\ref{46a}) in phase space with parameters as $k=1$, $M=1$.\\
ii)~~ If the isotropic pressure is less than the energy density then it has characteristics of ultra-relativistic fluid. In this type of fluid, the EoS takes the form $p=e/2$. After putting $k=1/2$ in Eq. (19) we obtain the following expression for the critical radius
\begin{eqnarray}\label{49}
r_c\simeq\frac{5}{2}GM-\frac{14}{25}\frac{\xi}{M}.
\end{eqnarray}
The Hamiltonian (\ref{41b}) in this case reduces to
\begin{eqnarray}\label{49b}
\mathcal{H}&=&\frac{\sqrt{\Big(1-\frac{2GM}{r}+\frac{2G^2M\xi}{r^3}\Big)}}{r^{2}v\sqrt{1-v^{2}}}.
\end{eqnarray}
We can observe that $\mathcal{H}$ in Eq. (\ref{49b}) is not defined for $(r,v^2)=(r_h,1)$. However, for some constant values of $\mathcal{H}=\mathcal{H}_0$ one can solve it for $v^2$. The five trajectories of solutions to (\ref{49b}) in phase space are shown in top right diagram of Figure 1. Here the black curve shows the solution for $\mathcal{H}=\mathcal{H}_{c}$, the red curve shows the solution for $\mathcal{H}=\mathcal{H}_{c}-0.01$, the green curve shows the solution for $\mathcal{H}=\mathcal{H}_{c}-0.005$, the magenta curve is for $\mathcal{H}=\mathcal{H}_{c}+0.01$ and the blue curve is for $\mathcal{H}=\mathcal{H}_{c}+0.005087$. Looking at the contour plots, we see that they are doubly valued and show unphysical behavior so we can say that there is no physical significance of such fluid in ASG.\\
iii)~~For the radiation fluid we have the state parameter $k=1/3$. This fluid has the property to absorbs the radiations emitted by the black hole. Insertion of $k=1/3$ in (19) results in the following real approximation of the critical radius
\begin{eqnarray}\label{51}
r_{c} &\simeq& 3GM-\frac{5}{9}\frac{\xi}{M},
\end{eqnarray}
while the Hamiltonian (21) takes the form
\begin{eqnarray}\label{52a}
\mathcal{H}&=&\frac{\Big(1-\frac{2GM}{r}+\frac{2G^2M\xi}{r^3}\Big)^{2/3}}{r^{4/3}v^{2/3}(1-v^{2})^{2/3}}.
\end{eqnarray}
From above Hamiltonian we see that the point $(r,v^2)=(r_h,1)$ is not a critical point of the dynamical Hamiltonian system. However, the expression for $v^2$ can be obtained by fixing $H=H_0$. The characteristics of solution curves are depicted in the left lower picture where the black curve shows the solution for $\mathcal{H}=\mathcal{H}_{c}$, the red curve shows the solution for $\mathcal{H}=\mathcal{H}_{c}+0.00099$, the green curve shows the solution for $\mathcal{H}=\mathcal{H}_{c}+0.0009$, the magenta curve is for $\mathcal{H}=\mathcal{H}_{c}-0.04$ and the blue curve is for $\mathcal{H}=\mathcal{H}_{c}-0.09$.\\
Here, for the radiation fluid we find some surprising characteristics as it gets closer to the black hole. The black, magenta and blue curves exhibit unphysical behavior, however, the green curves describes very interesting behavior of transonic type. Before the critical point the fluid has supersonic velocity but as soon as it approaches the critical point the speed becomes subsonic.\\
iv)~~In sub-relativistic fluids, energy density exceeds the isotropic pressure and the assigned value to the state parameter is $k=1/4$. Repeating the previous steps we obtain an approximation of the critical radius as
\be
r_c\simeq\frac{7}{2}GM-\frac{26}{49}\frac{\xi}{M}.
\ee
So insertion of (28) in (20) provides the desired critical point $(r_c,u_c)$.\\
In this case of sub-relativistic fluid, the Hamiltonian (24) takes the following form
\begin{eqnarray}\label{55a}
\mathcal{H}&=&\frac{\Big(1-\frac{2GM}{r}+\frac{2G^2M\xi}{r^3}\Big)^{3/4}}{rv^{1/2}(1-v^{2})^{3/4}}.
\end{eqnarray}
It is evident from above equation that the point $(r,v^2)=(r_h,1)$ is not a critical point of the dynamical system. Now we draw contour plots of $\mathcal{H}$ in the $(r,v)$ plane by fixing $\mathcal{H}=\mathcal{H}_c$ which describes the following behavior of the moving fluid. The black curve shows the solution for $\mathcal{H}=\mathcal{H}_{c}$, the red curve shows the solution for $\mathcal{H}=\mathcal{H}_{c}+0.03$, the green curve shows the solution for $\mathcal{H}=\mathcal{H}_{c}-0.0399$, the magenta curve is for $\mathcal{H}=\mathcal{H}_{c}+0.0009$ and the blue curve is for $\mathcal{H}=\mathcal{H}_{c}-0.0317999$. The curves shown in the lower right figure describe the behavior of the moving fluid as: in the blue and green curves, the fluid reach near the critical point to show transonic behavior but surprisingly failed to touch it. So in this scenario we define this motion as unphysical behavior of the fluid (as they show velocity as double valued function). However, blue and magenta curves show the supersonic accretion motion in the region $v>v_c$ and subsonic motion in the region where $v<v_c$.\\
%The plot shows the trajectories of solutions to Eq. (\ref{55a}) in phase space with parameters as $k=1/4$, $M=1$.
%\\
%In Fig.1 (top left) the fluid flow is transonic. The speed of flow increases as it reaches close to the black hole. In the top right figure, it is observed that the fluid bounces back before reaching the critical point. It is a sub-sonic flow. In the bottom left figure, similar behavior is observed. The speed of fluid is increased after the bounce. The bottom right figure shows both transonic and subsonic flows.
%
%In the figs shown by (\ref{f5}) we see the different behavior of fluids. For $k=1$ the fluid is supersonic for $v>0$ and subsonic for $v<0$. In the second figure for $k=1/2$ the fluid has no physical behavior. For $k=1/3$ and $k=1/4$ the fluid is again subsonic for $v<0$ and supersonic for $v>0$.

\subsection{Critical Analysis for the Isothermal Fluids}
Unlike the Schawrzschild black hole, the quantum gravity affects the accreting fluid near the improved version of Schawrzschild black hole. Furthermore, if we do not entertain the quantum gravity effects, the above presented results are easily reducible to what already been published in \cite{Michel}. Now we discuss the asymptotic behavior of isothermal fluids with EoS $p=ke$ such that $0<k<1$. It is easy to write the Eq. (13) in the following form
\begin{eqnarray}\label{new8}
\frac{GM}{2r_{c}}+\frac{3G^{2}M\xi}{2r_{c}^{3}}&=&k\Big(1-\frac{3GM}{2r_{c}^{2}}+\frac{7G^{2}M\xi}{2r_{c}^{3}}\Big),
\end{eqnarray}
which can again be reduced to a depressed cubic equation by introducing Wierstrass polynomial $r=t+\frac{7GM}{6}$ which is equivalently expressed as
\begin{eqnarray}\label{new9}
t_{c}^{3}-pt_c-q&=&0,
\end{eqnarray}
where
\begin{eqnarray}\label{new10}
p&=&\frac{49G^{2}M^{2}}{12},~~~~~q=\frac{343G^{3}M^{3}}{108}.
\end{eqnarray}
Here again the equation (\ref{new9}) has three roots: one real root and the other two will be complex conjugates of each other. This follows directly by the Cardano formula
\begin{eqnarray}\label{new11}
r_{c}&=&\sqrt[3]{-q+\iota\sqrt{W}}+\sqrt[3]{-q-\iota\sqrt{W}},
\end{eqnarray}
where $W=\sqrt{p^{3}+q^{2}}$. We can do a detailed analysis by computing the Jacobian matrix for the Eq. (10) and (11), for instance
\begin{equation}
J=
  \begin{pmatrix}
    \frac{\partial f_{1}}{\partial r} & \frac{\partial f_{1}}{\partial u} \\
    \frac{\partial f_{2}}{\partial r} & \frac{\partial f_{1}}{\partial u}
  \end{pmatrix}
  \label{new12}
\end{equation}
With the help of above Jacobian matrix one can determine that either the critical values are center, saddle or spiral. If both eigenvalues are real and have different signs we have a saddle point. If the real part of the complex eigenvalues is negative then we have a spiral and if the real part of the complex number is zero then we have a center.

\begin{figure}[!ht]
\centering
\includegraphics[width=15cm]{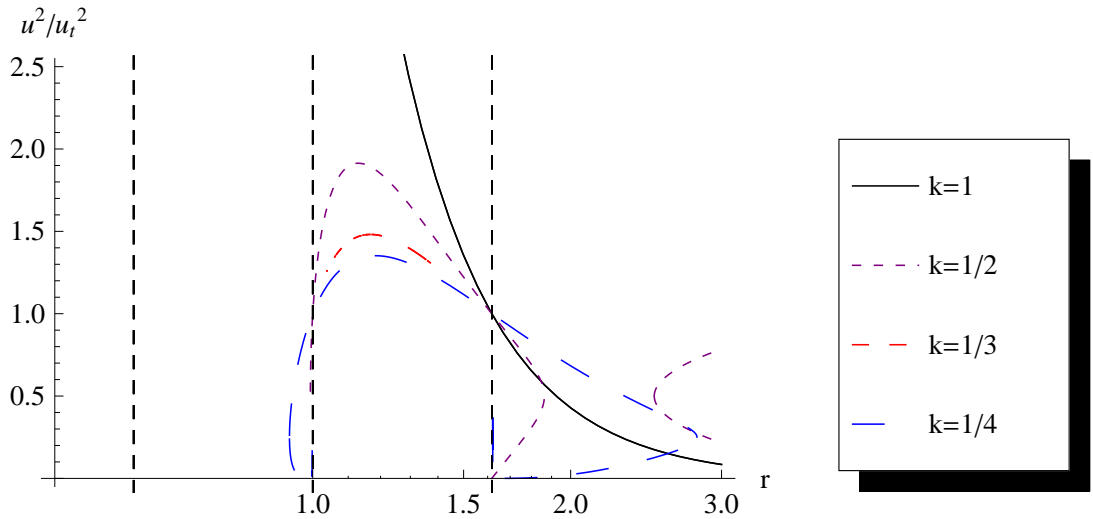}
\caption{The obtained transonic solutions for the isothermal fluid with equation of state $p=ke$ with $k=1, 1/2, 1/3, 1/4$. The value of coupling parameter $\xi=0.5$ and other constants are fixed to be $M=G=1$}\label{f}
\end{figure}
%\begin{figure}[!ht]
%	\centering
%	\includegraphics[width=10cm]{astt.eps}
%	\caption{The obtained transonic solutions for the isothermal fluid with equation of state $p=ke$ with $k=1, 1/2, 1/3, 1/4$. The value of coupling parameter $\xi=0.5$ and other constants are fixed to be $M=G=1$}\label{f}
%\end{figure}
So using Eqs. (19) and (20), one can obtain $r_c$ and $u_c^r$ respectively (velocity of the fluid at sonic point). Then, after putting $(r_c,u^r_c)$ in Eq. (14) we find the constant $A$ to obtain $u$ in an explicit form. Moreover, from the normalization of four-velocity vector one can also derive an expression for $u_t(r)$. So, after knowing the explicit forms of $u$ and $u_t$, one can sketch $\Big(\frac{u}{u_t}\Big)^2$ along $r-axis$ to see whether for each case k=1, 1/2, 1/3, 1/4, the fluid passes through the sonic point or not as sketched in Fig.2.
In Fig (\ref{f5}) we have discussed non-transonic solutions but we have also plotted the transonic solutions for the isothermal fluid in Fig.2. The transonic solutions yield maximum accretion rate because they pass through the critical point. In Fig.2, we see that the fluid trajectories may form an orbit for $k=1/4$ near the Cauchy and event horizon.

\section{Conclusion}
The model of spherical accretion is used generally to test various theories of modified gravity. To test these theories from the astrophysical perspective, one checks how does the behaviour of fluids modifies under the change of parameters of modified gravity appearing in the metric of black holes. By varying these free parameters of modified gravity, the positions of critical points might shift and the speed of fluid flow might enhance or decay near the black hole. Moreover, the fluid behavior might shift from supersonic to subsonic. In literature, the spherical accretion on black holes has been studied under the frameworks of different modified gravities like braneworld gravity \cite{Abb}, Horava-Lifshitz gravity \cite{Abdul}, f(R) gravity \cite{A1} and f(T) gravity \cite{A2}, to list a few. In the present paper, we are motivated to test another candidate theory of quantum gravity namely, the higher derivative asymptotic safe gravity in the infra-red limit. As can be seen from the Eq.3 of our paper that size of black holes in the ASG theory is smaller compared to Schwarzschild BH. Furthermore, Eq. 26 suggests that the position of critical point shifts more towards the BH and is smaller than the respective critical point for Schwarzschild BH. Thus one can compare and distinguish the relativistic accretion models from Schwarzschild BH from a ASG BH by changing the $\xi$ parameter.\\	
In this paper we adapt the Hamiltonian method of Michel type accretion as developed by some of the present authors \cite{A1}. This method is more general than the original method of Michel. Here we use the general equations for spherical accretion including conservation laws for the ASG BH static metric. The pressure of the perfect fluid for such spherically symmetric flows is, up to a sign, the Legendre transform of the energy density. This leads to a nice differential equation allowing the determination of the energy density, enthalpy, or pressure knowing one of the equations of state.
Furthermore, the Bondi's model of accretion on a normal star is the oldest model of spherical accretion using the Newtonian mechanics. In that model, the fluid is adiabatic and non-viscous and the flow is always transonic, thus it allows the existence of critical point. The Bondi model also allows the outflows during accretion which can explain the jet phenomenon from certain active galactic galaxies, see for details \cite{Shub}. However, we found that the fluid flow can be more general than Bondi model. The fluid flow can have subsonic, supersonic and transonic regimes. Also due to relativistic treatment, more than one critical points might exist which allow the heteroclinic flows as well.\\
In our paper, we have studied both adiabatic and isothermal fluid flows since both of them have important astrophysical relevance, see \cite{Shub}. In other words, there are astrophysical situations where either entropy is constant and temperature varying (adiabatic) or vice versa (isothermal). In the isothermal case, the sound speed of accretion flow at any radii is always equivalent to the sound speed at sonic point. Hence, if the temperature of the flow is known, one can easily compute the critical point.\\
In the Newtonian stellar accretion model, the size of critical radius is considerably larger than the relevant
	Schwarzschild radius of the star (several hundreds or thousands times the Schwarzschild radius of the star), while the corresponding critical radius for the ASG improved black hole is between 2M and 6M or comparable to three times the Schwarzschild radius. Therefore for accretion over black holes, the fluid experiences transonic or ultra-sonic flow just seconds before entering the horizon. The ASG black holes are predicted to be smaller in size compared to the Schwarzschild black hole, while
	both having same mass. Thus the relevant transitions from sonic to supersonic to ultrasonic flows occur at much faster rate for ASG black holes. In the study of accretion near the black hole, we have found that there is no physical significance of the radiation fluid exist in asymptotic safe gravity. We have observed that the effect of ASG parameter affects the fluid behavior for small values of radial parameter. Moreover, in the critical analysis of the isothermal test fluids we have seen that the fluid trajectories may form a closed orbit for $k=1/4$ near the Cauchy horizon. It is interesting to consider astronomical observation effects, such as accretion rate and temperature, as done in \cite{Yang, yang1, Frank}, which are our future researches.\\
\textbf{Acknowledgment:} The authors would like to thank anonymous referee for providing insightful comments.   

\end{document}